\documentstyle[aps,epsfig,amssymb]{revtex}

\begin{document}

\draft
\title{Magnetic Quantum Oscillations of the Longitudinal Conductivity $\sigma_{zz}$ in Quasi two-dimensional Metals}
\author{T. Champel and V.P. Mineev}
\address{
Commissariat \`{a} l'Energie Atomique,  DSM/DRFMC/SPSMS\\
17 rue des Martyrs, 38054 Grenoble Cedex 9, France}
\date{\today}

\maketitle

\begin{abstract}
We derive an analytical expression for 
the longitudinal magnetoconductivity $\sigma_{zz}$ in layered conductors in presence of a quantizing magnetic field perpendicular to the layers and for short-range in-plane impurity scattering
in frame of the quantum transport theory.
% in the regime $\mu \gg \omega_{c}$, where $\mu$ is the chemical potential 
%and $\omega_{c}$ is the cyclotron pulsation
%An internal reservoir, formed by the quasi-one dimensional part of 
%Fermi surface present in many quasi two-dimensional organic compounds, and 
%having influence on the impurity self-energy through its associated density 
%of states is taken into account in the self-consistent Born approximation.
Our derivation points out quite unusual temperature and magnetic field dependences 
for Shubnikov-de Haas oscillations
 in the two-dimensional limit, i.e. $\hbar \omega_{c} \gg 4 \pi t$, where $t$ is the interlayer hopping integral for electrons, and $\omega_{c}$ the cyclotron frequency.
In particular, when  $\hbar \omega_{c} \gg 4 \pi t$ and $\hbar \omega_{c} \geq 2 \pi \Gamma_{\mu}$ (here $\Gamma_{\mu}$ is the value of the imaginary part of the impurity self-energy at the chemical potential $\mu$), a pseudo-gap centered on integer values of $\mu/\hbar\omega_{c}$ appears in the zero-temperature magnetoconductivity function $\sigma_{zz}(\mu/\hbar\omega_{c})$.
At low temperatures, this high-field regime is characterized by a thermally activated behavior of the conductivity minima (when chemical potential $\mu$ lies between Landau levels) in correspondence with the recent observation in the organic conductor $\beta''\text{-(BEDT-TTF)}_{2}\text{SF}_{5}\text{CH}_{2}\text{CF}_{2}\text{SO}_{3}$.

\end{abstract}


\section{Introduction}

In the past few years, numerous studies of the effects of a quantizing magnetic field in low-dimensional systems, especially in quasi two-dimensional (2D) organic conductors have been performed (see e.g.~\cite{1} and references therein).
The quasi-2D conductors 
consist typically
of highly
conducting planes with a small interplane quasiparticle dispersion.
In presence of a magnetic field $H$ perpendicular to the layers, the motion in the conducting planes is quantized.
The quasiparticule dispersion is then written as
\begin{equation}
\epsilon_{n}(p_{z})=(n+1/2) \omega_{c}-2t\cos p_{z}s
\end{equation}
where $s$ is the interlayer distance, $t$ the interplane transfer integral, $p_{z}$ the interlayer momentum in $z$ direction, $n$ a positive integer, $\omega_{c}=eH/m^{\ast}$ the cyclotron frequency depending on the quasiparticle effective mass $m^{\ast}$ and $e$ the electron's charge (Planck's constant, Boltzmann's constant and the velocity of light are chosen to be unity throughout the paper).
The Landau quantization of the 2D motion gives rise to the well-known magnetic quantum oscillations effects such as de Haas-van Alphen (dHvA) effect - magnetization oscillations - or Shubnikov-de Haas (SdH) effect - magnetoresistance oscillations.
In quasi-2D organic conductors, the resistivity component in the direction perpendicular to the layers $\rho_{zz}$ is  typically studied~\cite{1}, when both electrical current and magnetic field are parallel to same direction $z$.

Recently, J. Wosnitza {\em et al.}~\cite{2} and M.-S. Nam {\em et al.}~\cite{3} reported unusually strong oscillations of $\rho_{zz}$ in the quasi-2D organic conductor $\beta''\text{-(BEDT-TTF)}_{2}\text{SF}_{5}\text{CH}_{2}\text{CF}_{2}\text{SO}_{3}$ in high-magnetic fields at low temperatures.
The authors of Ref.~\cite{2} have called their observation a field-induced metal-insulator transition. M.-S. Nam {\em et al.}~\cite{3} proposed that 
the insulating-like behavior stems from the presence of a gap in the density of states by pointing out the thermally activated behavior of maxima of $\rho_{zz}$ (or minima of conductivity $\sigma_{zz}=\rho_{zz}^{-1}$), which occur when the chemical potential $\mu$ lies between Landau levels.
Their interpretation is based on the existence of negligibly small chemical potential oscillations with magnetic field owing to the quasi-1D part of Fermi surface which plays the role of an infinite reservoir of states.
The assumption of a fixed chemical potential is consistent with the information provided
by the inverse sawtooth wave form of the magnetization oscillations observed in the same compound~\cite{2,4}, which is a strong indication of negligibly small chemical potential oscillations (see e.g.~\cite{5} and references therein).

Some articles~\cite{6,7,8,9,10} are devoted to the theoretical study of the oscillations of magnetoconductivity $\sigma_{zz}$ in quasi-2D metals.
Very recently,  P.D. Grigoriev {\em et al.} have proposed in a series of papers~\cite{7,8,9} an explanation of other features concerning $\sigma_{zz}$ oscillations such as
the field-dependent phase shift of the beats and the slow oscillations of $\sigma_{zz}$, in frame of both the semiclassical Boltzmann theory~\cite{7} and the quantum transport theory~\cite{8,9}.
They showed that these features observed in layered organic conductors other than  $\beta''\text{-(BEDT-TTF)}_{2}\text{SF}_{5}\text{CH}_{2}\text{CF}_{2}\text{SO}_{3}$ originate from the pronounced low-dimensionality of energy spectrum (1) in the low-field limit  $\omega_{c} \leq 4 \pi t$.
It is worth noting that the latter condition is not fulfilled in the range of magnetic fields where experiments of Ref.~\cite{2,3} were done. The observation of very sharp extrema in the dHvA oscillations~\cite{2,4} hints that the 2D limit $4 \pi t \ll \omega_{c}$ is most probably reached~\cite{5}.

In this paper,
we 
investigate theoretically the magnetoconductivity oscillations
with a view to give an explanation to the experiments reported by J. Wosnitza {\em et al.}~\cite{2} and M.-S. Nam {\em et al.}~\cite{3}.
General calculations of SdH oscillations of magnetoconductivity $\sigma_{zz}$ in quasi-2D metals are performed in frame of the quantum transport theory for short-range impurity scattering.
In the second section, the general expression for $\sigma_{zz}$ is represented in a more convenient form valid for $\mu \gg \omega_{c}$ and an arbitrary self-energy.  We demonstrate in Section III the existence of a thermally activated behavior for the minima of magnetoconductivity in high quality quasi-2D metals in high magnetic fields. This behavior is shown to be not related to the presence of a gap in the density of states but rather to the presence of a pseudo-gap in the zero-temperature magnetoconductivity function $\sigma_{zz}(\mu/\omega_{c})$.
In Section IV, the impurity self-energy for the layered quasi-2D metals is investigated in frame of the self-consistent Born approximation.
Finally, in Section V, our results are discussed in light of the experimental observations in $\beta''\text{-(BEDT-TTF)}_{2}\text{SF}_{5}\text{CH}_{2}\text{CF}_{2}\text{SO}_{3}$~\cite{2,3}.
The detailed description of the studied microscopic model and the corresponding transport calculations are given in Appendix.

\section{The longitudinal conductivity $\sigma_{zz}$}

In absence of vertex corrections, the longitudinal conductivity $\sigma_{zz}$ in quasi-2D metals under a quantizing magnetic field perpendicular to the layers can be expressed as (see Appendix)

\begin{equation}
\sigma_{zz}=e^{2}s \frac{g_{0}}{2} \omega_{c}\sum_{n,\sigma}\int\frac{dp_{z}}{2\pi}v_{z}^{2}(p_{z})\int \frac{d\varepsilon}{2\pi}\left(-n'_{F}(\varepsilon)\right)\left(G^{R}_{\sigma,n,p_{z},\varepsilon}G^{A}_{\sigma,n,p_{z}, \varepsilon}
-\Re \, \left(G^{R}_{\sigma,n,p_{z},\varepsilon}\right)^{2}
\right),
\end{equation}
where summation has to be done over the energy $\varepsilon$ and the magnetic quantum numbers $n$ of the Landau levels, the spin projections $\sigma=\pm 1$, and the interlayer momentum $p_{z}$.
Here  $\Re$ means the real part, $g_{0}$ is the 2D density of states in absence of magnetic field, $n'_{F}(\varepsilon)$ is the derivative of the Fermi-Dirac distribution function
\begin{equation}
n'_{F}(\varepsilon)=-\frac{1}{4T \cosh^{2}(\varepsilon/2T)},
\end{equation}
and $T$ the temperature.
The limits of the integral over $p_{z}$ are $[-\pi/s, + \pi/s]$, and those of the integral over $\varepsilon$ are $[-\infty, + \infty]$.
The functions $G^{R}$ and $G^{A}$ are the retarded and advanced Green's functions defined by
\begin{equation}
G_{\sigma,n,p_{z}, \varepsilon}^{A}=\left(G_{\sigma,n,p_{z}, \varepsilon}^{R}\right)^{\ast}=\frac{1}{\varepsilon-\xi_{\sigma,n, p_{z}}+ i \Gamma_{\varepsilon+\mu}},
\end{equation}
where
 \begin{equation}
\xi_{\sigma,n,p_{z}}=\left(n+\frac{1}{2}\right) \omega_{c}-2t \cos p_{z}s-\mu-\sigma \mu_{e} H,
\end{equation}
 $\mu_{e}$ is the electron's magnetic moment,
and $\Sigma(\varepsilon)=i\Gamma_{\varepsilon}$ is the energy-dependent impurity self-energy. The real part of self-energy $\Sigma$ is included in chemical potential $\mu$.
In our model, $\Sigma$ is independent of the quantum numbers $n,p_{z},\sigma$ because we consider only point-like impurity scattering.
Lastly,
 $v_{z}$ is the velocity in the $z$ direction
\begin{equation}
v_{z}(p_{z})=\frac{\partial \xi_{\sigma,n, p_{z}}}{\partial p_{z}}=2st\sin p_{z}s.
\end{equation}
Formula (2) is one half smaller than that used by Grigoriev {\em et al.}~\cite{9} without derivation. 

In Eq. (2), two terms contribute to $\sigma_{zz}$ : one originating from the product of the retarded and the advanced Green's functions (which we call $RA$), the other from the product of two retarded (or two advanced) Green's functions (contribution $RR$).
Further, we choose to calculate the two terms $RR$ and $RA$ separately in order to see clearly how they contribute in the final expression:
\begin{equation}
\sigma_{RA}=e^{2}s \frac{g_{0}}{2} \omega_{c}\sum_{n,\sigma}\int\frac{dp_{z}}{2\pi}v_{z}^{2}(p_{z})\int \frac{d\varepsilon}{2 \pi}(-n'_{F}(\varepsilon))G^{R}_{\sigma,n,p_{z},\varepsilon}G^{A}_{\sigma,n,p_{z}, \varepsilon}
\end{equation}
and 
\begin{equation}
\sigma_{RR}=e^{2}s \frac{g_{0}}{2} \omega_{c}\sum_{n,\sigma}\int\frac{dp_{z}}{2\pi}v_{z}^{2}(p_{z})\int \frac{d\varepsilon}{2 \pi}n'_{F}(\varepsilon)
\Re
\left(G^{R}_{\sigma,n,p_{z},\varepsilon}\right)^{2}.
\end{equation}
In contrast with the case of metals without magnetic field  where the contribution $RR$ is absent (see e.g. B.L. Altshuler {\em et al.}~\cite{11}), we show in the third section that the term $RR$ contributes here to the total conductivity, and specifically in the 2D limit.

Now, we rewrite expressions (7) and (8) by performing the summation over the Landau levels $n$.
For convenience, we neglect henceforward the spin splitting, which can be easily introduced in the following calculations.
To transform the sum over Landau levels, we use the Poisson summation formula
\begin{equation}
\sum_{n=0}^{+\infty}f(n)=\sum_{l=-\infty}^{+\infty} \int_{a}^{+ \infty}du f(u) e^{-2 \pi ilu},
\end{equation}
where $a$ is any number between $-1$ and $0$,
to obtain for the first contribution $RA$
\begin{eqnarray}
\sigma_{RA}=e^{2}s g_{0} \omega_{c}
\sum_{l=-\infty}^{+\infty}\int 
\frac{dp_{z}}{2 \pi}
v^{2}_{z}(p_{z})
\int\!\!d \varepsilon
\left(-n'_{F}(\varepsilon)\right)
\int \frac{du}{2\pi} \,
e^{-2 \pi ilu} \times \nonumber
\\
\times
\frac{1}{\left(\varepsilon-\xi_{u,p_{z}}-i\Gamma_{\varepsilon+\mu}\right)\left( \varepsilon-\xi_{u,p_{z}}+i\Gamma_{\varepsilon+\mu}\right)}.
\end{eqnarray}
Changing the variable $u$ to $x=\xi_{u,p_{z}}-\varepsilon$, and then setting the lower limit of the integral over variable $x$ to $-\infty$ since we consider the relevant regime $\mu \gg \omega_{c}$ for SdH oscillations, we get

\begin{eqnarray}
\sigma_{RA}&=&e^{2}sg_{0} \sum_{l=-\infty}^{+\infty}\left(-1\right)^{l}
\int \frac{dp_{z}}{2 \pi}
v^{2}_{z}(p_{z})
e^{-2 \pi il\frac{2t}{\omega_{c}} \cos p_{z}s}
\int \!\!d \varepsilon
\left(-n'_{F}(\varepsilon)\right)
\,e^{-2 \pi il \frac{\mu+\varepsilon}{\omega_{c}}}
\times \nonumber \\
&&\times 
 \int \!\! \frac{dx}{2\pi} \,e^{-2 \pi il\frac{x}{\omega_{c}}}
\frac{1}{\left(x- i \Gamma_{\varepsilon+\mu}\right)\left(
x+ i\Gamma_{\varepsilon+\mu}\right)}.
\end{eqnarray}
A straigthforward residue calculation for the integral over $x$ yields

\begin{eqnarray}
\sigma_{RA}=e^{2}s\frac{g_{0}}{2}\sum_{l =-\infty}^{+ \infty}\left(-1\right)^{l}\int
\frac{dp_{z}}{2\pi}v^{2}_{z}(p_{z})e^{-2 \pi i l \frac{2t}{\omega_{c}} \cos p_{z}s}  
\int \!\! d \varepsilon
\frac{\left(-n_{F}'(\varepsilon)\right)}{\Gamma_{\varepsilon+\mu}}
 e^{-2 \pi i l \frac{\varepsilon+\mu}{\omega_{c}}}
e^{-2 \pi |l| \frac{\Gamma_{\varepsilon+\mu}}{\omega_{c}}}.
\end{eqnarray}
Collecting positive and negative $l$, and introducing the  first order Bessel function $J_{1}(x)$ defined by
$$
\int_{-\pi}^{+\pi}\frac{d \varphi}{2 \pi} \sin^{2} \varphi e^{-ix  \cos \varphi}=\frac{J_{1}(x)}{x},$$
we rewrite Eq. (12) as

\begin{eqnarray}
\sigma_{RA}=\sigma_{0}
\int \!\!
d \varepsilon \left(-n_{F}'(\varepsilon) \right)
\frac{\Gamma_{0}}{\Gamma_{\varepsilon+\mu}}
\left[1+
\frac{\omega_{c}}{\pi t}
\sum_{l=1}^{+\infty} \frac{\left(-1\right)^{l}}{l} J_{1} \left(\frac{4 \pi lt}{\omega_{c}}\right)
e^{-2 \pi l \frac{\Gamma_{\varepsilon+\mu}}{\omega_{c}}} \cos\left(2 \pi l \frac{\varepsilon+\mu}{\omega_{c}}\right)\right],
\end{eqnarray}
where 
\begin{equation}
\sigma_{0}=e^{2}g_{0}\frac{\left(ts\right)^{2}}{\Gamma_{0}}=e^{2}g_{0}\langle v_{z}^{2}\rangle \tau_{0}
\end{equation}
is the longitudinal conductivity in absence of magnetic field, $\Gamma_{0}=1/2 \tau_{0}$ the imaginary part of the impurity self-energy at zero magnetic field, $\tau_{0}$ the mean free time of quasiparticles also at zero magnetic field.
As for $\sigma_{RR}$, we perform the sum over the Landau levels in the same way to get  
\begin{eqnarray}
\sigma_{RR}=\sigma_{0}
\int \!\!d \varepsilon \left(-n_{F}'(\varepsilon) \right)
\frac{2 \Gamma_{0}}{t}
\sum_{l=1}^{+\infty} \left(-1\right)^{l} J_{1} \left(\frac{4 \pi lt}{\omega_{c}}\right)
e^{-2 \pi l \frac{\Gamma_{\varepsilon+\mu}}{\omega_{c}}} \cos\left(2 \pi l \frac{\varepsilon+\mu}{\omega_{c}}\right).
\end{eqnarray}

Let's compare both contributions $RA$ and $RR$.
If we consider the limit of zero magnetic field, the contribution $\sigma_{RR}$ vanishes and we recover the usual Drude formula given by
\begin{equation}
\sigma_{zz} = \sigma_{RA}=\sigma_{0}.
\end{equation}
We find again that in absence of magnetic field the contribution $RA$ corresponds to the semi-classical contribution to transport~\cite{11}. The term $RR$ has no semi-classical equivalent. 
In the 3D limit $4 \pi t \gg \omega_{c}$ or in the dirty limit $2\pi \Gamma_{\varepsilon} \gg \omega_{c}$, the contribution $RR$ represents actually a small correction compared to main contribution $RA$. However,
at arbitrary magnetic field, the comparison of the two contributions is not trivial, since both terms are expressed as alternated series.
We demonstrate in next section that the contribution $RR$ becomes quite significant specifically in the two-dimensional limit when $4 \pi t \ll \omega_{c}$.

\section{Magnetoconductivity in the two-dimensional limit}

In
 the 2D limit $\omega_{c} \gg 4 \pi t$, the two contributions $\sigma_{RA}$ and $\sigma_{RR}$ (given by Eq. (13) and (15)) get simplified since the Bessel function factors can be expanded as
$
J_{1}(x) \sim x/2$
for small values of argument $x$:
\begin{equation}
\sigma_{RA}=\sigma_{0}
\int \!\! d \varepsilon \left(-n_{F}'(\varepsilon) \right)
\frac{\Gamma_{0}}{\Gamma_{\varepsilon+\mu}}
\left[
1+2
\sum_{l=1}^{+\infty} \left(-1\right)^{l}
e^{-2 \pi l \frac{\Gamma_{\varepsilon+\mu}}{\omega_{c}}}
\cos\left(2 \pi l \frac{\varepsilon+\mu}{\omega_{c}}\right)
\right]
\end{equation}
and 
\begin{equation}
\sigma_{RR}=\sigma_{0}
\int \!\!d \varepsilon \left(-n_{F}'(\varepsilon) \right)
\frac{4\pi\Gamma_{0}}{\omega_{c}}
\sum_{l=1}^{+\infty} l \left(-1\right)^{l}
e^{-2 \pi l \frac{\Gamma_{\varepsilon+\mu}}{\omega_{c}}}
\cos\left(2 \pi l \frac{\varepsilon+\mu}{\omega_{c}}\right).
\end{equation}
In this limit, the field oscillating part of $\sigma_{zz}$ becomes independent of the interlayer hopping integral $t$, which appears only in the zero-field normalization factor $\sigma_{0}$. This clearly establishes the physical origin of $\sigma_{zz}$ oscillations as the quantization  of motion in the plane perpendicular to the magnetic field, which affects the interlayer transport through energy spectrum (1).

In order to compare both contributions $\sigma_{RA}$ and $\sigma_{RR}$ at arbitrary magnetic field such that $\omega_{c} \gg 4 \pi t$, summation over the integers $l$ is performed for each contribution.
Then, the total magnetoconductivity is written as
\begin{equation}
\sigma_{zz}=\int_{-\infty}^{+\infty}d \varepsilon \left(-n_{F}'(\varepsilon-\mu) \right)\left[\sigma_{RA}(\varepsilon)+\sigma_{RR}(\varepsilon)\right],
\end{equation}
where
\begin{equation}
\sigma_{RA}(\varepsilon)=\sigma_{0}\frac{\Gamma_{0}}{\Gamma_{\varepsilon}} 
\frac{
\sinh\left(2 \pi \frac{\Gamma_{\varepsilon}}{\omega_{c}}\right)}
{\cosh\left(2 \pi \frac{\Gamma_{\varepsilon}}{\omega_{c}}\right)+
\cos\left(2 \pi \frac{\varepsilon}{\omega_{c}}\right)}
\end{equation}
and
\begin{equation}
\sigma_{RR}(\varepsilon)=-\sigma_{0}
\frac{2\pi \Gamma_{0}}{\omega_{c}}
\frac{
1+\cosh\left(2 \pi \frac{\Gamma_{\varepsilon}}{\omega_{c}}\right)\cos\left(2 \pi \frac{\varepsilon}{\omega_{c}}\right)}
{\left(\cosh\left(2 \pi \frac{\Gamma_{\varepsilon}}{\omega_{c}}\right)+
\cos\left(2 \pi \frac{\varepsilon}{\omega_{c}}\right)\right)^{2}}.
\end{equation}
At zero temperature, magnetoconductivity oscillations are then given by
\begin{equation}
\sigma_{zz}=\sigma_{RA}(\mu)+\sigma_{RR}(\mu)=\sigma_{zz}(\mu).
\end{equation}
As already noted at the end of Section II, the term $\sigma_{RR}$ contributes negligibly to the total conductivity $\sigma_{zz}$ in the dirty limit $2 \pi \Gamma_{\varepsilon} \gg \omega_{c}$.
From Eq. (21) we can see that at zero temperature $\sigma_{RR}$ gives always a negative contribution to $\sigma_{zz}(\mu)$ when $\mu \sim n \omega_{c}$, and a positive contribution when $\mu \sim (n+1/2) \omega_{c}$. 

When $\omega_{c} \gg 2 \pi \Gamma_{\varepsilon}$, $\omega_{c} \gg 4 \pi t$ and $\mu \neq (n+1/2) \omega_{c}$, both contributions $RA$ and $RR$ are of the same order so that the total magnetoconductivity at zero temperature vanishes periodically, $\sigma_{zz}(\mu) \rightarrow 0$.
Such a limit indicates the existence of a particular behavior for the form of SdH oscillations at low temperatures, which is the subject of the following development.

Let's consider that the oscillations of self-energy are negligibly small, i.e.
 $\Gamma_{\varepsilon} \approx \Gamma_{0}$. This assumption is not a necessary ingredient for our theory. We only use it to study  the behavior of $\sigma_{zz}$ in high magnetic fields on a more quantitative level. However, the relation $\Gamma_{\varepsilon} \approx \Gamma_{0}$ is proved to be correct in particular in the self-consistent Born approximation in presence of an internal reservoir of states in the conditions of experiment~\cite{3} (see next Section).

\begin{figure}
\begin{centering}
\epsfig{file=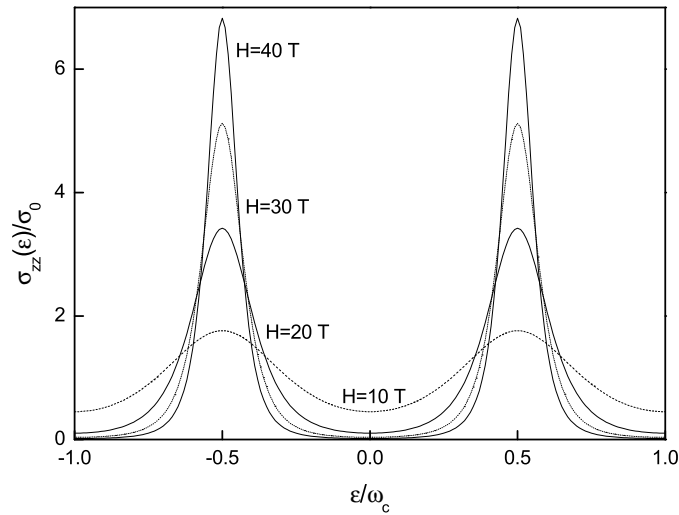}
\caption{Presence of a pseudo-gap in the function $\sigma_{zz}(\varepsilon)$ for high magnetic fields. The opening of a pseudo-gap occurs for  $20$ T, i.e. for a parameter $\alpha=2 \pi \Gamma_{0}/\omega_{c} \approx 1$. Thus, the other magnetic fields $H=$ 10, 30 and 40 T correspond respectively to $\alpha=$ 2, 2/3 and 0.5.
} 
\label{fig1}
\end{centering}
\end{figure}

The numerical analysis of the function $\sigma_{zz}(\varepsilon)$ reveals two regimes depending on the ratio $\alpha=2\pi\Gamma_{0}/\omega_{c}$ for the shape of oscillations:
\begin{itemize}
\item
A low-field regime $\omega_{c} < 2 \pi \Gamma_{0}$ (or $\alpha >1$), where the contribution $RR$ represents a correction of the $RA$ contribution for the whole range of energy $\varepsilon$,
\item
A high-field regime defined by $\omega_{c} > 2 \pi \Gamma_{0}$ (or $\alpha <1$), where a pseudo-gap centered on the energy $\varepsilon=n\omega_{c}$ exists in the function $\sigma_{zz}(\varepsilon)$.
\end{itemize}
In
Fig. \ref{fig1} we represent the function $\sigma_{zz}(\varepsilon)$ for different values of magnetic fields, using the experimental values given in Ref.~\cite{3}, namely  $\Gamma_{0}=$0.22 meV and $m^{\ast}=1.96 m_{e}$ (here $m_{e}$ is the electron mass). 
As shown in Fig. \ref{fig1}, the opening 
of a pseudo-gap occurs for a magnetic field of about 20 T, corresponding to a parameter $\alpha \sim 1$. The width of the pseudo-gap clearly increases with the magnetic field.

Such a high-field behavior can also be found from analytical considerations.
Indeed, 
the presence of a pseudo-gap in  $\sigma_{zz}(\varepsilon)$  is clear if we write the total conductivity as
\begin{equation}
\sigma_{zz}(\varepsilon)=\sigma_{0}\left[
\frac{\sinh \alpha-\alpha \cos \left(2 \pi \frac{\varepsilon}{\omega_{c}}\right)}
{\cosh \alpha+\cos\left(2 \pi \frac{\varepsilon}{\omega_{c}}\right)}
-\alpha \frac{\sin^{2}\left(2 \pi \frac{\varepsilon}{\omega_{c}}\right)}
{\left(\cosh \alpha+\cos\left(2 \pi \frac{\varepsilon}{\omega_{c}}\right)\right)^{2}}
\right].
\end{equation}
For $\varepsilon \sim n \omega_{c}$ and $\alpha <1$, magnetoconductivity $\sigma_{zz}(\varepsilon)$ becomes very small over a range in energy, which gets more and more wide with a decreasing $\alpha$.

\begin{figure}
\begin{centering}
\epsfig{file=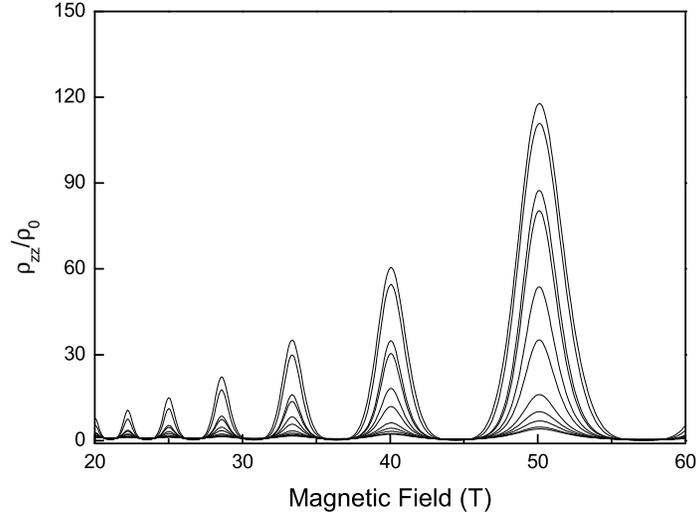}
\caption{Oscillations of the magnetoresistance $\rho_{zz}/\rho_{0}=\sigma_{0}/\sigma_{zz}$ as a function of magnetic field for different temperatures (from the top, 0.59, 0.94, 1.48, 1.58, 1.91, 2.18, 2.68, 3.03, 3.38, 3.80, and 4.00 K).} 
\label{fig2}
\end{centering}
\end{figure}

Such small values for $\sigma_{zz}$ when $\alpha <1$ lead to huge maxima of magnetoresistance $\rho_{zz}$ 
at low temperatures when $\mu/\omega_{c} \sim n$.
To illustrate the behavior at finite temperature, we have calculated numerically $\sigma_{zz}$ using Eq. (19)-(21) still taking experimental values~\cite{3}. We estimated the chemical potential at $\mu \approx 11.8$ mev (for the numerical study, we have explicitly assumed that the chemical potential oscillations are negligibly small; effects of chemical potential oscillations are evoked in the dicussion).
In Fig. \ref{fig2}, the
general form of $\rho_{zz}=\sigma_{zz}^{-1}$ oscillations for negligibly small self-energy oscillations is represented in the range of magnetic fields 20 - 60 T. In this plot, we have chosen exactly the same temperatures as in Fig. 1 of Ref.~\cite{3} in order to compare more easily the shapes of magnetoresistance oscillations.

\begin{figure}
\begin{centering}
\epsfig{file=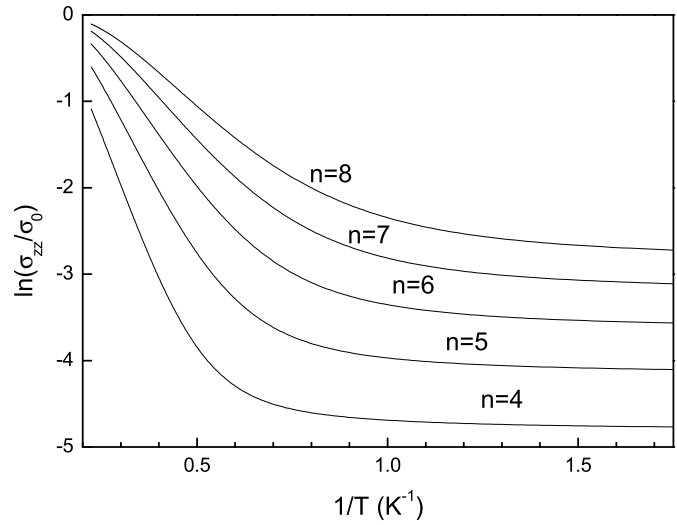}
\caption{Plot of $\ln \sigma_{zz}/\sigma_{0}$ versus $1/T$ ($T$ in Kelvins) for the integers $n=\mu/ \omega_{c}$ (maxima of resistivity).
} 
\label{fig3}
\end{centering}
\end{figure}

As in Ref.~\cite{3}, we have also studied numerically the temperature dependence of the resistivity maxima. The presence of a pseudo-gap for high-magnetic fields gives rise 
 at finite temperature to a thermally activated behavior of conductivity minima at $\mu \sim n \omega_{c}$. 
Indeed, according to Eq. (19) and when $\mu \sim n \omega_{c}$ (then the minima of $\sigma_{zz}(\varepsilon)$ are in correspondence with the maximum of $-n'_{F}(\varepsilon)$), the main means of conduction is the thermal excitation of quasiparticles at the edges of the pseudo-gap.
Figure \ref{fig3} shows $\ln \sigma_{zz}$ versus $1/T$ for the integer values $\mu/\omega_{c}=n=$4, 5, 6, 7 and 8.
In this plot, a linear dependence is clearly found for the interval of temperatures $2 <T < 4$ K (or $0.25 < T^{-1} < 0.5 \text{ K}^{-1}$), pointing out the activated behavior of $\sigma_{zz}$ minima.
At very low temperatures $T < 2$ K (or $T^{-1} > 0.5 \text{ K}^{-1}$),  $\sigma_{zz}$ saturates (Fig. \ref{fig3}): quasiparticles whose energy lies in the pseudo-gap mainly contribute to conduction.

\section{The impurity Self-energy}
For the numerical study of preceding section, we have assumed that the oscillations of the self-energy are negligibly small in the 2D limit $\omega_{c} \gg 4 \pi t$. The purpose of this section is to justify such a choice in the organic compound $\beta''\text{-(BEDT-TTF)}_{2}\text{SF}_{5}\text{CH}_{2}\text{CF}_{2}\text{SO}_{3}$.
The impurity self-energy is derived in the so-called self-consistent Born approximation.
In quasi-2D metals, we usually have to take into account of the presence of an internal reservoir of states~\cite{6,4}.
For short-range impurity scatterers, the imaginary part $\Gamma$ of the self-energy is directly proportional to the total density of states $g^{tot}$, which is taken as the sum of the quasi-2D density of states $\tilde{g}$ and of the field-independent reservoir's density of states $g^{R}$. We assume that the same relation of proportionality holds both in presence and in absence of magnetic field and thus
\begin{equation}
\frac{\Gamma_{\varepsilon}}{\Gamma_{0}}= \frac{g^{tot}(\varepsilon)}{g^{tot}_{0}}
=\frac{1}{1+R}\left(R+\frac{\tilde{g}(\varepsilon)}{\tilde{g}_{0}} \right),
\end{equation}
where $R=g^{R}/\tilde{g}_{0}$ is a parameter measuring the strength of the reservoir, $\tilde{g}(\varepsilon)$ the density of states of the quasi-2D metal in presence of a quantizing magnetic field, and $\tilde{g}_{0}=g_{0}/s$ the quasi-2D density of states at zero magnetic field.
For simplicity, we neglect the dependence of the reservoir's density of states $g^{R}$ on energy.
The quasi-2D density of states $\tilde{g}(\varepsilon)$ depends on the impurity self-energy through the expression
\begin{equation}
\tilde{g}(\varepsilon)=\frac{i}{2 \pi}\sum_{\sigma,n,p_{z}}
\left(G^{R}_{\sigma,n,p_{z},\varepsilon-\mu}-G^{A}_{\sigma,n,p_{z},\varepsilon-\mu}\right),
\end{equation}
which for $\mu \gg \omega_{c}$ can also be written (without spin-splitting) after application of the Poisson summation formula

\begin{equation}
\tilde{g}(\varepsilon)=\tilde{g}_{0}\left[1+2\sum_{l=1}^{+\infty}\int s\frac{dp_{z}}{2\pi}
\left(-1\right)^{l}
e^{-2 \pi l \frac{\Gamma_{\varepsilon}}{\omega_{c}}}
\cos\left( \frac{2\pi l}{\omega_{c}}\left(\varepsilon+2t\cos p_{z}s \right)
\right)
\right].
\end{equation}
Then, after summation over momentum $p_{z}$ we obtain~\cite{12} 
\begin{equation}
\tilde{g}(\varepsilon)=\tilde{g}_{0}\left[1+2\sum_{l=1}^{+\infty}
\left(-1\right)^{l}
J_{0}\left(\frac{4\pi l t}{\omega_{c}}\right)
e^{-2 \pi l \frac{\Gamma_{\varepsilon}}{\omega_{c}}}
\cos\left(2\pi l \frac{\varepsilon}{\omega_{c}}\right)
\right],
\end{equation}
where $J_{0}$ is the zero order Bessel function.
Combination of Eq. (24) and Eq. (27) yields a complicated self-consistent equation to solve for the impurity self-energy $\Gamma_{\varepsilon}$.

For  $4 \pi t \geq \omega_{c}$, it is reasonable to keep only the first term $l=1$ in the sum of Eq. (27). As a result, the impurity self-energy oscillates weakly with $H$. In absence of reservoir ($R=0$), this field dependence has to be taken into account in the expressions (13) and (15) of the longitudinal conductivity $\sigma_{zz}$.
 This regime has already been investigated by Grigoriev {\em et al.}~\cite{7,8,9} who showed that the field-dependent phase shift of the beats and the slow oscillations of $\sigma_{zz}$ are consequently 
general features of the SdH oscillations in low-dimensional metals.

In general, we have to take into account many terms in the sum of Eq. (27). 
By summing over the integers $l$ before summing over momentum $p_{z}$ in Eq. (26), we can write the quasi-2D density of states $\tilde{g}(\varepsilon)$ in a more convenient form

\begin{equation}
\tilde{g}(\varepsilon)=\tilde{g}_{0}  \int_{-\pi}^{+ \pi}\frac{d \varphi}{2 \pi} \frac{\sinh\left(2\pi\frac{\Gamma_{\varepsilon}}{\omega_{c}}\right)}{\cosh\left(2\pi\frac{\Gamma_{\varepsilon}}{\omega_{c}}\right)+\cos\left(2\pi\frac{\varepsilon}{\omega_{c}}+\frac{4\pi t}{\omega_{c}}\cos \varphi \right)},
\end{equation}
so that the self-consistent equation to solve in quasi-2D metals becomes
\begin{equation}
\frac{\Gamma_{\varepsilon}}{\Gamma_{0}}= 
\frac{1}{1+R}\left(R+
 \int \frac{d \varphi}{2 \pi} \frac{\sinh\left(2\pi\frac{\Gamma_{\varepsilon}}{\omega_{c}}\right)}{\cosh\left(2\pi\frac{\Gamma_{\varepsilon}}{\omega_{c}}\right)+\cos\left(2\pi\frac{\varepsilon}{\omega_{c}}+\frac{4\pi t}{\omega_{c}}\cos \varphi \right)}
\right).
\end{equation}
For $\omega_{c} \ll 2 \pi \Gamma_{0}$, the Landau levels are strongly broadened by impurity scattering and the oscillations of impurity self-energy scattering  are negligibly small; this is not the interesting regime. Expression (29) enables to study the impurity self-energy at arbitrary magnetic fields.
The self-consistent equation to solve in the 2D limit $\omega_{c} \gg 4 \pi t$
 takes the form
\begin{equation}
\frac{\Gamma_{\varepsilon}}{\Gamma_{0}}=\frac{R}{1+R} +\frac{1}{1+R}\frac{\sinh\left(2\pi\frac{\Gamma_{\varepsilon}}{\omega_{c}}\right)}{\cosh\left(2 \pi \frac{\Gamma_{\varepsilon}}{\omega_{c}}\right)+\cos\left(2\pi\frac{\varepsilon}{\omega_{c}}\right)}.
\end{equation}

 This equation in absence of reservoir ($R=0$) is similar to that obtained many years ago by T. Ando for a 2D metal~\cite{13}. For magnetic fields $\omega_{c} \geq \pi \Gamma_{0}$, it leads formally to a zero value for the self-energy $\Gamma_{\varepsilon}$ when energy $\varepsilon$ lies between two adjacent Landau levels~\cite{13}. But for $\Gamma_{\varepsilon}=0$, one can not fulfill the summation in (26) (taken with $t=0$ for the 2D metal). This contradiction demonstrates that the self-consistent Born approximation is inappopriate in the high magnetic field region $\omega_{c} \geq \pi \Gamma_{0}$.
To avoid this unphysical result obtained for a short-range impurity scattering, M.E. Raikh and T.V. Shahbazyan~\cite{14} (see also Ref.~\cite{15}) have discussed the 2D electron system in a random potential with a finite correlation radius $R_{c}$. It was shown that the self-consistent Born approximation is justified only in the low field region when the magnetic length $l_{H} \ll R_{c}$. At high magnetic fields such that
 $R_{c} \geq l_{H}$, the higher order 
diagrams contributions become important and have also to be taken into account in the impurity self-energy calculations. As a result, the density of states is proved to be finite at any energy and for arbitrary magnetic fields.

However, for $R \neq 0$, we note that the unphysical cut-off for the impurity self-energy does not exist.
This can be shown by rewriting Eq. (30) as
\begin{equation}
\cos\left(2\pi\frac{\varepsilon}{\omega_{c}}\right)=
\frac{\sinh\left(2\pi\frac{\Gamma_{\varepsilon}}{\omega_{c}}\right)}
{(1+R)\Gamma_{\varepsilon}/\Gamma_{0}-R}
-\cosh\left(2 \pi \frac{\Gamma_{\varepsilon}}{\omega_{c}}\right).
\end{equation}
This equation admits solutions for the whole range of energy $\varepsilon$ and for any magnetic fields
when
$\Gamma_{\varepsilon} > \Gamma_{0}R/(1+R)$. Thus, it is always possible to find a finite impurity self-energy at arbitrary energy and magnetic fields.
For this reason, we consider in the present paper that in the 2D limit the oscillations of $\Gamma_{\varepsilon}$ are given by Eq. (31). The validity of the self-consistent Born approximation in the 2D limit and in presence of a reservoir is addressed to future work.

\begin{figure}
\begin{centering}
\epsfig{file=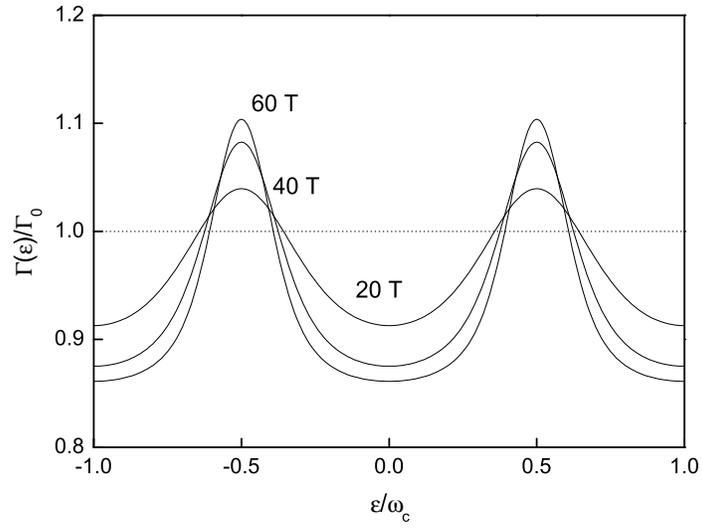}
\caption{Oscillations of the impurity self-energy $\Gamma_{\varepsilon}$ in the 2D limit for different magnetic fields (20, 40 and 60 T). Here $R=5$, $\Gamma_{0}=0.22$ meV and $m^{\ast}=1.96 m_{e}$.}
\label{fig4}
\end{centering}
\end{figure}

We plot in Fig. \ref{fig4} the oscillations of $\Gamma_{\varepsilon}$ for different magnetic fields $H=$20, 40 and 60 T, and for a reservoir parameter $R=5$ (the choice of such a value was suggested by Ref.~\cite{4}), using the same parameters as in Fig.~\ref{fig1}.
As expected, the oscillations amplitude increases with the magnetic field (see Fig. \ref{fig4}).
However, for $R=5$, oscillations of self-energy $\Gamma_{\varepsilon}$ are relatively small compared to $\Gamma_{0}$ even at magnetic fields of about 60 T.
In absence of other informations about the importance of the reservoir's influence on self-energy oscillations, we consequently assume at this level that oscillations of $\Gamma_{\varepsilon}$ are small in experiments of Ref.~\cite{2,3}.

\section{Discussion}

Numerical treatment of Eq. (19)-(21) allows  comparison with experiments~\cite{2,3}.
Theoretical calculations with negligibly small self-energy oscillations reproduce well the appearance of huge oscillations in the same conditions of temperature and magnetic fields as in experiments~\cite{2,3}.
As demonstrated in third section, the thermally activated behavior of $\rho_{zz}$ maxima is theoretically established in correspondence with the observations of Ref.~\cite{3}. Furthermore, the minimum value of magnetic field to have thermal activation, i.e. $2 \pi \Gamma_{0} \sim \omega_{c}$, is in quantitative agreement with experiment (in the conditions of experiment~\cite{3}, it corresponds to a magnetic field of about 20 T).

Nevertheless, we notice that there exist some discrepancies between our calculations and experiment~\cite{2,3} especially at very low temperatures.
Indeed, our numerical calculations do not qualitatively reproduce the behavior at very low temperatures of the resistivity minima, which are increasing with the magnetic field (compare Fig.~\ref{fig2} of the present paper and Fig. 1 of Ref.~\cite{3}).
Quantitative departures from experiment are also found for the resistivity maxima at very low temperatures.
A reason for the latter may be the consideration of negligibly small self-energy oscillations in our calculations. Indeed, although the presence of self-energy oscillations does not suppress the thermally activated behavior 
of
$\rho_{zz}$ maxima, it induces quantitative modifications for the pseudo-gap value and thus for the resistivity maxima.
However, this can not qualitatively explain the field dependence of the resistivity minima at very low temperatures.

A possible limitation of our model to describe the experimental signal~\cite{2,3} is the consideration of point-like impurity scattering, which enables to neglect the vertex corrections. 
Indeed,
starting from a microscopic model with the energy spectrum (1) (i.e. a 3D model), we find that in presence of point-like impurity scattering the behavior of the magnetic quantum oscillations of $\sigma_{zz}$ is fully dictated for magnetic fields $\omega_{c} \gg 4\pi t$ by the 2D in-plane motion. 
Under the same condition and in a parallel direction, the equation for the density of states of the quasi-2D metal investigated with the self-consistent Born approximation takes also a 2D form.
On the other hand, in the pure 2D case,
 the treatment of the impurity effects on the magnetic quantum oscillations by a short-range impurity scattering is known to be no more valid at high magnetic fields to describe accurately the density of states.
According to Ref.~\cite{14}, when typically the magnetic length $l_{H} \leq R_{c}$, it becomes necessary to include also the effects of the impurity scattering on finite range $R_{c}$ in the calculations of the density of states.
Probably,
specific contributions due to the finite-range impurity scattering have also to be taken into account in the calculations of the magnetoconductivity oscillations in high magnetic fields.
In our opinion, such an open question might explain why our present calculations do not seize the complete correct behavior of magnetic quantum oscillations of $\rho_{zz}$ in the high field regime.

So far, we did not discuss the possible effects of the chemical potential oscillations on the magnetoconductivity oscillations. As declared in introduction, the dHvA signal~\cite{4} gives strong support for negligibly small chemical potential oscillations in the organic compound $\beta''\text{-(BEDT-TTF)}_{2}\text{SF}_{5}\text{CH}_{2}\text{CF}_{2}\text{SO}_{3}$, what we use at the final stage to perform the numerical calculations.
The mechanism responsible for fixing the chemical potential is not clearly established. Actually, this question is directly related to the theoretical problem just evoked above of the density of states of a 2D metal in high quantizing magnetic fields.
Nevertheless, it is interesting to note that the value of the pseudo-gap appearing in the zero-temperature magnetic field-dependence of $\sigma_{zz}$ depends strongly on the presence of chemical potential oscillations. In third section, it was shown that in the limit $\omega_{c} \rightarrow \infty$ the magnetoconductivity $\sigma_{zz}(\mu)$ vanishes everywhere but at $\mu = (n+1/2)\omega_{c}$.
In a pure metal in absence of reservoir, the chemical potential is pinned to the Landau levels, i.e. $\mu=(n+1/2)\omega_{c}$ when varying the magnetic field. Therefore $\sigma_{zz}$ does not vanish, except when the chemical potential jumps.
Thus, in the pure (unrealistic) case, the pseudo-gap in $\sigma_{zz}$ is reduced to a point and there is no thermally activated behavior for the conductivity minima.
In a real system, the chemical potential oscillations at zero temperature lead to a smaller pseudo-gap value than that obtained at a constant chemical potential.
 Although no full quantitative agreement between our numerical calculations and experiment has been found for the pseudo-gap value,
the observation~\cite{3} of a well developed thermally activated regime with a relatively important pseudo-gap value seems accordingly consistent with the existence of negligibly small chemical potential oscillations.

We point out that the observation of a drastic change in the temperature and magnetic field behavior of $\sigma_{zz}$ oscillations is a unique feature of the 2D limit with energy spectrum (1). Since the switch between the two regimes occurs for $\omega_{c} \sim 2 \pi \Gamma_{0}$, this necessarily implies that $\Gamma_{0} > 2t$.
Such a condition is exactly expected to be reached in the organic compound $\beta''\text{-(BEDT-TTF)}_{2}\text{SF}_{5}\text{CH}_{2}\text{CF}_{2}\text{SO}_{3}$~\cite{2}.
It is worth noting that the same condition is proposed for having incoherent interlayer transport~\cite{2}. In that case, the existence of an interlayer dispersion in the energy spectrum (1), or in other words the Fermi liquid description, is questioned. It was with such an idea that authors of Ref.~\cite{2} attempted to extract a background magnetoresistance superposed with SdH oscillations.
In a recent experiment~\cite{16} performed with the same organic compound as concerned here, another arguments in favor of incoherent transport have been proposed to interpret observations of the angular dependence of $\rho_{zz}$ in high magnetic fields.
However, we notice that the coherent magnetotransport in quasi-2D compounds is usually discussed in frame of the Boltzmann transport theory.
Here, we clearly find important discrepancies between the results obtained with the quantum transport theory and those obtained with the Boltzmann transport theory especially in high magnetic fields $\omega_{c} \gg 4\pi t$ and $\omega_{c} \geq \Gamma_{0}$. 
Therefore, theoretical studies developed exclusively in frame of the quantum transport theory are needed to better understand the possible unconventional quantum features of coherent magnetotransport in low-dimensional systems.
As for the possible absence of the quasi-particle description at $t < \tau_{0}^{-1}$ in the quasi-2D metals~\cite{16}, one can notice that there is no doubt concerning the quasi-particle concepts in the pure 2D case where $t=0$, and the only criterion for this rests on the Fermi energy: $\varepsilon_{F} > \tau_{0}^{-1}$. It seems that the possibility of an interlayer motion at $t\neq 0$ does not spoil the quasi-particle description because the in-plane scattering just leads to the undetermination  $\delta k \sim \left(v_{F}\tau_{0}\right)^{-1}$ 
of the in-plane momentum ($v_{F}$ is the velocity at the Fermi surface) but not to the indetermination of the interplane quasi-momentum $k_{z}$. This explains in our opinion the observations of the coherence in the quasi-2D organic conductor $\kappa \text{-(BEDT-TTF)}_{2}\text{Cu}\text{(NCS)}_{2}$
 reported at $ \tau_{0}^{-1} \sim 6 t$ in Ref.~\cite{17}.

Finally, the importance of the contribution $\sigma_{RR}$ in the 2D limit and in high-quality layered conductors demonstrated in Section III prevents from separating the magnetoconductivity $\sigma_{zz}$ into a field oscillating part and a steady part in the manner used by the authors of Ref.~\cite{2}.
The relation of proportionality between the derivative of dHvA signal and SdH signal assumed in Ref.~\cite{2} breaks down in high magnetic fields and in low-dimensional systems. 
 This already occurs for $\omega_{c} \leq 4 \pi t$ due to small self-energy oscillations in absence of reservoir~\cite{7,8,9}.
Furthermore, even in presence of negligibly small self-energy oscillations, the relation of proportionality may not be valid.
In this latter case
the integration over energy $\varepsilon$ in (17) can be performed analytically and yields for the oscillating part of the contribution $RA$ in the 2D limit: 
\begin{eqnarray}
\sigma_{RA}-\sigma_{0}&=&2\sigma_{0}
\sum_{l=1}^{+\infty} \left(-1\right)^{l}
e^{-2 \pi l \frac{\Gamma_{0}}{\omega_{c}}}
\cos\left(2 \pi l \frac{\mu}{\omega_{c}}\right)
\int \!\! d \varepsilon \left(-n_{F}'(\varepsilon) \right)
\cos\left(2 \pi l \frac{\varepsilon}{\omega_{c}}\right) \\
&=&
2\sigma_{0}
\sum_{l=1}^{+\infty} \left(-1\right)^{l}
e^{-2 \pi l \frac{\Gamma_{0}}{\omega_{c}}}
\cos\left(2 \pi l \frac{\mu}{\omega_{c}}\right)
 \frac{\lambda_{l}}
{\sinh \lambda_{l}},
\end{eqnarray}
where we have introduced
\begin{equation}
\lambda_{l}=\frac{2\pi^{2}lT}{\omega_{c}}.
\end{equation}
In quasi-2D metals, the magnetization oscillations $M$ 
are given in the 2D limit by~\cite{12}
\begin{equation}
M=g_{0}\mu 
\frac{\omega_{c}}{H}
\frac{1}{\pi} 
\sum_{l=1}^{+\infty} \frac{\left(-1\right)^{l+1}}{l}
 \frac{\lambda_{l}}
{\sinh \lambda_{l}}
e^{-2 \pi l \frac{\Gamma_{0}}{\omega_{c}}} \sin \left(2 \pi l \frac{\mu}{\omega_{c}}\right) .
\end{equation}
From Eq. (33) and (35), 
we can derive (in the case of negligibly small chemical potential oscillations) a relation valid at any temperature between the derivative of $M$ and the oscillating part of the contribution $RA$ by keeping only the rapidly oscillating terms:
\begin{equation}
\frac{\sigma_{RA}-\sigma_{0}}{\sigma_{0}}
\approx
\frac{H^{2}}{g_{0}\mu^{2}} \,\, \frac{dM}{dH}.
\end{equation}
In the dirty limit $\omega_{c} \ll 2 \pi \Gamma_{0}$, it has been shown in section II that $\sigma_{RR}$ contributes negligibly to total conductivity, i.e. $\sigma_{zz} \approx \sigma_{RA}$, so that the SdH oscillations are approximatively proportional to $H^{2}dM/dH$, exactly as in 3D metals.
When $\omega_{c} \geq 2 \pi \Gamma_{0}$, the contribution $\sigma_{RR}$ becomes quite important and there is then no more simple correspondence between dHvA and SdH signals.

\section{Conclusion}

We have investigated the oscillations of the longitudinal magnetoconductivity in quasi-2D organic metals for impurity scattering on short-range in a perpendicular quantizing magnetic field. 
Huge magnetoresistance oscillations  with unconventional magnetic field and temperature dependences
 have been found in high magnetic fields and for low temperatures in layered conductors whose interlayer hopping integral $4 \pi t \ll \omega_{c}$.
In particular, we showed that
 the activated behavior of resistivity maxima recently observed in the organic conductor
$\beta''\text{-(BEDT-TTF)}_{2}\text{SF}_{5}\text{CH}_{2}\text{CF}_{2}\text{SO}_{3}$~\cite{3}
is a general feature of Shubnikov-de Haas oscillations in the 2D limit $\omega_{c} \gg 4 \pi t$ in good samples and/or in very high magnetic fields $\omega_{c} \geq 2 \pi \Gamma_{0}$.
The origin of such a behavior is not related to the presence of a gap in the density of states but rather to the presence of a pseudo-gap in the zero temperature magnetoconductivity function $\sigma_{zz}(\mu/\omega_{c})$.

\appendix
\section*{Calculation of $\sigma_{zz}$}

The longitudinal conductivity $\sigma_{zz}$ characterizes the appearence of a current density in the direction $z$ in response to an external electrical field $E_{z}$
\begin{equation}
j_{z}=\sigma_{zz}E_{z}.
\end{equation}
With the objective to find the longitudinal conductivity, an alternative monochromatic vector potential ${\cal A}_{z}$ depending on the Matsubara variable $\tau$  is formally introduced in the form
\begin{equation}
{\cal A}_{z}(\vec{r},j,\tau)={\cal A}_{z}(\vec{k},k_{z},\omega)e^{i\vec{k}\cdot \vec{r}+i k_{z}sj-i \omega \tau}
\end{equation}
and such that
\begin{equation}
E_{z}=-\frac{\partial {\cal A}_{z}}{\partial t}.
\end{equation}
Here, $j$ is an integer which indexes the layer and $\vec{r}=(x,y)$ is the position in the layer.
The relation between $E_{z}$ and ${\cal A}_{z}$ takes a simpler form in Fourier space 
\begin{equation}
E_{z}(\vec{k},k_{z}, \omega)=i \omega{\cal A}_{z}(\vec{k},k_{z}, \omega).
\end{equation}
Then, the relation between the formal field ${\cal A}_{z}$ and the current density in Fourier representation is
\begin{equation}
j_{z}(\vec{k},k_{z}, \omega)=s \sum_{j}\int \!\!\!
\int j_{z}(\vec{r},j,\tau)e^{-i\vec{k} \cdot \vec{r}-ik_{z}sj+i\omega \tau}
d^{2}r \,d\tau=\sigma_{zz}i \omega {\cal A}_{z}(\vec{k},k_{z}, \omega)
\end{equation}

\subsection{Hamiltonian}
We consider the same model Hamiltonian as that used in the paper of R.A. Klemm {\em et al.}~\cite{18}, in which the electrons are free to propagate within each layer, and tunnel between the layers (in the vertical direction $z$).
The Hamiltonian is constructed in the second quantization language, where the field operator $\Psi_{j, \sigma}(\vec{r})$ corresponding to an electron with spin  $\sigma$ at position $\vec{r}$ in the $j$th layer is defined by
\begin{equation}
\Psi_{j,\sigma}(\vec{r})=\frac{1}{\sqrt{Ss}} \sum_{\vec{k}}e^{i \vec{k} \cdot \vec{r}}a_{j, \sigma}(\vec{k}),
\end{equation}
where $a_{j,\sigma}(\vec{k})$ is the annihilation operator and $S$ is the area of a layer ($S=1$ after).
The Hamiltonian we shall consider is of the form
\begin{equation}
{\cal H}={\cal H}_{0}+{\cal H}_{T}+{\cal H}_{imp}
\end{equation}
where ${\cal H}_{0}$ is the two-dimensional Hamiltonian for an electron in a magnetic field $\vec{H}=H \vec{z}$ associated with a vector potential $\vec{A}$
\begin{equation}
{\cal H}_{0}=s \int d^{2}r 
\sum_{j, \sigma}
\left[
\Psi_{j, \sigma}^{\dagger}
(\vec{r})
\left(\frac{1}{2m^{\ast}}\left(-i\vec{\nabla}_{\vec{r}}-e\vec{A}\right)^{2}-\mu_{e} \vec{\sigma} \cdot \vec{H} \right)\Psi_{j,\sigma}(\vec{r})\right],
\end{equation}
${\cal H}_{T}$ is the gauge-invariant tunneling Hamiltonian, which models the coupling between the layers :
\begin{equation}
{\cal H}_{T}=-t s \int d^{2}r \sum_{j, \sigma} \Psi_{j,\sigma}^{\dagger}(\vec{r})\Psi_{j+1, \sigma}(\vec{r}) \exp\left(-ie\int_{js}^{(j+1)s}{\cal A}_{z}(\vec{r},j,\tau)dz\right) +H.C.
\end{equation}
Here the expression $H.C.$ means the Hermitian conjugate.
Finally, ${\cal H}_{imp}$ describes the coupling of electrons with the impurities in the layers (we neglect the scattering that might occur in the tunneling process)
\begin{equation}
{\cal H}_{imp}=s \int d^{2}r\sum_{j,\sigma} \Psi_{j,\sigma}^{\dagger}(\vec{r})u_{j}(\vec{r})\Psi_{j, \sigma}(\vec{r}),
\end{equation}
where $u_{j}(\vec{r})$ is the short-range scattering potential.

% le potentiel de diffusion sur les impuret\'{e}s  est donn\'{e} par
%\begin{equation}
%V_{jl}(\vec{r}, \vec{r}')=d^{-1} \delta_{jl} \sum_{s} \int\int \frac{d^{2} 
%p \, d^{2}q}{ (2 \pi)^{4}} e^{i\vec{p} \cdot \left(\frac{1}{2}(\vec{r}+
%\vec{r}')-\vec{R}_{s}^{j}\right)+i\vec{q} \cdot (\vec{r}-\vec{r}')}\left(V_{1}%+iV_{SO}\,\hat{p} \times \hat{q} \cdot \sigma \right)
%\end{equation}
%o\`{u} $R_{s}^{j}$ est la position de la $s$i\`{e}me impuret\'{e} dans la 
%couche $j$, $\hat{p}$ et $\hat{q}$ sont des vecteurs unitaires sur la surface %de Fermi, et $V_{1}$ et $V_{SO}$ sont respectivement les contributions 
%ind\'{e}pendantes du spin et de type spin-orbite de la diffusion.

In our model, the longitudinal current density $j_{z}$ originates only from the tunneling process, and is then given by
\begin{equation}
\hat{j_{z}}(\vec{r},j,\tau)=-\frac{\delta {\cal H}_{T}}{\delta {\cal A}_{z}(\vec{r},j,\tau)}
\end{equation}
 The derivation of the Hamiltonian ${\cal H}_{T}$ yields for a small external field ${\cal A}_{z}$
\begin{eqnarray}
\hat{j_{z}}(\vec{r},j, \tau)&=&-iets^{2} \sum_{\sigma}
\left[\Psi_{j,\sigma}^{\dagger}(\vec{r})\Psi_{j+1, \sigma}(\vec{r}) \exp\left(-ise{\cal A}_{z}(\vec{r},j,\tau)\right) -H.C.\right] \\
&\approx& 
-iets^{2}
\sum_{\sigma} 
\left[
\Psi_{j,\sigma}^{\dagger}(\vec{r})\Psi_{j+1, \sigma}(\vec{r})
-\Psi_{j+1, \sigma}^{\dagger}(\vec{r})\Psi_{j,\sigma}(\vec{r})-
\right.
\nonumber \\ 
&&\left.
-\left(\Psi_{j,\sigma}^{\dagger}(\vec{r})\Psi_{j+1, \sigma}(\vec{r}) 
+\Psi_{j+1, \sigma}^{\dagger}(\vec{r})\Psi_{j,\sigma}(\vec{r})\right)
ies {\cal A}_{z}(\vec{r},j,\tau)\right]
\end{eqnarray}

\subsection{Green's functions}
To proceed further, we shall use the usual temperature Green's functions defined by~\cite{19}
\begin{equation}
G_{i,j}^{\sigma_{1},\sigma_{2}}(\vec{r}_{1},\tau_{1},\vec{r}_{2}, \tau_{2})=-\left\langle T_{\tau} \tilde{\Psi}_{i,\sigma_{1}}(\vec{r}_{1}, \tau_{1})\tilde{\Psi}^{\dagger}_{j,\sigma_{2}}(\vec{r}_{2}, \tau_{2}) \right \rangle
\end{equation}
where the averages are evaluated in the grand canonical ensemble, and $\tilde{\Psi}_{i,\sigma_{1}}(\vec{r}_{1}, \tau_{1})$ are the Heisenberg field operators depending on the Matsubara variable $\tau_{1}$.
Therefore, writing the averaged current density in the $j$th layer in terms of the Green's function of the system, we obtain
\begin{equation}
j_{j}(\vec{r},\tau)=\langle \hat{j_{z}}(\vec{r},j,\tau)\rangle=j_{j}^{P}(\vec{r},\tau)+j_{j}^{D}(\vec{r},\tau)
\end{equation}
where 
\begin{equation}
j_{j}^{P}(\vec{r},\tau)=-iets^{2}
\sum_{\sigma} 
\left(
G^{\sigma,\sigma}_{j+1, j}(\vec{r},\tau,\vec{r},\tau)-G^{\sigma,\sigma}_{j, j+1}(\vec{r},\tau,\vec{r},\tau)\right)
\end{equation}
is the paramagnetic part of the current density and
\begin{equation}
j_{j}^{D}(\vec{r},\tau)=-e^{2}ts^{3}
{\cal A}_{z}(\vec{r},j,\tau)
\sum_{\sigma} 
\left(
G^{\sigma,\sigma}_{j, j+1}(\vec{r},\tau,\vec{r},\tau)+G^{\sigma,\sigma}_{j+1, j}(\vec{r},\tau,\vec{r},\tau)\right)
\end{equation}
is the diamagnetic part of the current density.

The equation of motion for the Green's function  in the presence of impurities, magnetic field, and the vector potential ${\cal A}_{z}(\vec{r},j,\tau)$ is, for a given spin $\sigma$
\begin{eqnarray}
\left[
-\frac{\partial}{\partial \tau} 
-\frac{1}{2m^{\ast}}\left(-i\vec{\nabla}_{\vec{r}}-e\vec{A}\right)^{2}
+t\left(\hat{\Delta}_{j+}e^{-ies{\cal A}_{z}}+\hat{\Delta}_{j-}e^{ies {\cal A}_{z}}\right)
+\mu+\mu_{e} \sigma H- 
\right.
\nonumber \\
\left.
-u_{j}(\vec{r})
\right]
G_{\sigma, j,l}(\vec{r},\tau,\vec{r}',\tau')
=s^{-1}\delta_{jl}\delta(\vec{r}-\vec{r}')\delta(\tau-\tau')
\end{eqnarray}
where the operators $\hat{\Delta}_{j \pm}$ are defined by
\begin{equation}
\hat{\Delta}_{j \pm} \Psi_{j, \sigma}(\vec{r},\tau)=\Psi_{j\pm 1, \sigma}(\vec{r},\tau).
\end{equation}
The average of Eq. (A18) over impurities configuration is done using $\langle u_{j}(\vec{r})u_{l}(\vec{r}')\rangle=u_{0}^{2}s^{-1}\delta_{jl}\delta(\vec{r}-\vec{r}')$ (here $u_{0}$ is the amplitude of impurity potential). Keeping for the averaged Green's function $\langle G \rangle$  the same notation $G$ as before, we get for the equation of motion for the averaged Green's function 
\begin{eqnarray}
\left[
-\frac{\partial}{\partial \tau} 
-\frac{1}{2m^{\ast}}\left(-i\vec{\nabla}_{\vec{r}}-e\vec{A}\right)^{2}
+t\left(\hat{\Delta}_{j+}e^{-ies{\cal A}_{z}}+\hat{\Delta}_{j-}e^{ies {\cal A}_{z}}\right)
+\mu+\mu_{e} \sigma H+
\right.
\nonumber
\\
\left.
+\Sigma
\right]
G_{\sigma, j,l}(\vec{r},\tau,\vec{r}',\tau')
=s^{-1}\delta_{jl}\delta(\vec{r}-\vec{r}')\delta(\tau-\tau')
\end{eqnarray}
where 
$\Sigma$ is the impurity self-energy.

Next, we write the Green's function $G_{j,l}$ in the form
\begin{equation}
G_{\sigma,j,l}(\vec{r},\tau, \vec{r}', \tau')=G_{\sigma,j-l}^{0}(\vec{r},\vec{r}',\tau-\tau')+G_{\sigma,j,l}^{(1)}(\vec{r}, \tau, \vec{r}', \tau')
\end{equation}
where $G_{j, l}^{0}$ is the Green's function in the absence of ${\cal A}_{z}$ and $G_{j,l}^{(1)}$ is the first correction which is linear in ${\cal A}_{z}$.

It is more convenient to write $G^{0}$ in the Landau representation : 

\begin{eqnarray}
G_{\sigma,j-l}^{0}(\vec{r},\vec{r}',\tau-\tau')=
T\sum_{\omega_{\nu}}
e^{-i\omega_{\nu}(\tau-\tau')}
\int_{-\pi/s}^{\pi/s}\frac{dp_{z}}{2 \pi} 
e^{ip_{z}s(j-l)}
\int \frac{dp_{y}}{2 \pi} e^{ip_{y}(y-y')} \times
\nonumber
\\
\times
\sum_{n}
\Psi_{n}\left(x-p_{y}l_{H}^{2}\right)
\Psi_{n}\left(x'-p_{y}l_{H}^{2}\right)
G^{0}_{\sigma,n, p_{z}}(\omega_{\nu})
\end{eqnarray}
where
\begin{equation}
G^{0}_{\sigma,n, p_{z}}(\omega_{\nu})=\frac{1}{i \omega_{\nu}-\xi_{\sigma,n,p_{z}}+ \Sigma(\omega_{\nu})}
\end{equation}
and
\begin{equation}
\Psi_{n}(x)=\frac{1}{\left(\pi l_{H}^{2}\right)^{1/4}} \frac{1}{2^{n/2}\sqrt{n!}}\exp\left(-\frac{x^{2}}{2l_{H}^{2}}\right)H_{n}\left(\frac{x}{l_{H}}\right).
\end{equation}
Here $H_{n}$ is the Hermitte's polynom of degree $n$, $l_{H}=1/\sqrt{eH}$ is the magnetic length,
\begin{equation}
\xi_{\sigma,n,p_{z}}=\left(n+\frac{1}{2}\right) \omega_{c}-2t \cos p_{z}s-\mu-\sigma \mu_{e} H,
\end{equation}
$\omega_{\nu}=\pi T (2 \nu+1)$ are Matsubara frequencies and $\Gamma(\omega_{\nu})$ is the imaginary part of the impurity self-energy $\Sigma$. The real part of $\Sigma$ is included in chemical potential.
Here we consider that $\Sigma$ is independent of the quantum numbers, which is valid for short-range impurity scattering.

As for $G^{(1)}$, it is given by
\begin{equation}
G^{(1)}_{\sigma,j,l}(\vec{r},\tau,\vec{r'},\tau')=s
\sum_{k}\int\!\!\!\! \int \!\!d^{2}r'' d\tau''
G^{0}_{ \sigma, j-k}(\vec{r},\vec{r}'', \tau-\tau'')
V_{k}(\vec{r}'',\tau'')G^{0}_{ \sigma, k-l}(\vec{r}'',\vec{r}', \tau''-\tau')
\end{equation}
where $V_{k}$ is the potential
\begin{equation}
V_{k}(\vec{r},\tau)=iets{\cal A}_{z}(\vec{r},k,\tau)\left(\hat{\Delta}_{k+}-\hat{\Delta}_{k-}\right)
\end{equation}
 which is taken as a perturbation.

\subsection{Diamagnetic current}
At first order in the vector potential ${\cal A}_{z}$, the diamagnetic current is thus
\begin{equation}
j_{j}^{D}(\vec{r},\tau)=-e^{2}ts^{3}
{\cal A}_{z}(\vec{r},j,\tau)
\sum_{\sigma}\left(
G^{0}_{\sigma,-1}(\vec{r},\vec{r},0)+G^{0}_{\sigma,1}(\vec{r},\vec{r},0) 
\right)
\end{equation}
From Eq. (A22), we have straightforwardly
\begin{equation}
G^{0}_{\sigma,j-l}(\vec{r},\vec{r},0)=\frac{g_{0}}{2} \omega_{c} T \sum_{\omega_{\nu}}
\int_{-\pi/s}^{\pi/s}\frac{dp_{z}}{2 \pi} 
e^{ip_{z}s(j-l)}
\sum_{n}
G^{0}_{\sigma,n, p_{z}}(\omega_{\nu})
\end{equation}
where 
\begin{equation}
\frac{g_{0}}{2}\omega_{c}=\frac{1}{2 \pi l_{H}^{2}}=\int \frac{dp_{y}}{2 \pi} \left[\Psi_{n}\left(x-p_{y}l_{H}^{2}\right)\right]^{2}
\end{equation}
is the degeneracy of Landau level per spin state.
Then, the diamagnetic current is
\begin{equation}
j^{D}_{j}(\vec{r},\tau)=-e^{2}ts^{3} {\cal A}_{z}(\vec{r},j,\tau) g_{0} \omega_{c} \sum_{n,\sigma}\int \frac{dp_{z}}{2 \pi}\cos (p_{z}s) T \sum_{\omega_{\nu}} G^{0}_{\sigma,n,p_{z}}(\omega_{\nu}).
\end{equation}

Next, we transform the summation over Matsubara frequencies into an integration, and perform the analytical continuation~\cite{20}
\begin{equation}
T \sum_{\omega_{\nu}}G_{\sigma,n,p_{z}}(\omega_{\nu})=\frac{i}{4 \pi} \int_{-\infty}^{+\infty}d\varepsilon f_{\varepsilon} \left(G^{A}_{\sigma,n,p_{z},\varepsilon}-G^{R}_{\sigma,n,p_{z},\varepsilon}\right)
\end{equation}
where
\begin{equation}
f_{\varepsilon}=\tanh \frac{\varepsilon}{2 T}=1-2n_{F}(\varepsilon).
\end{equation}
Then, rewriting the expression for the diamagnetic current density by integrating by parts over the variable $p_{z}$, and making the Fourier transformation, we find
\begin{equation}
j^{D}(\vec{k},k_{z},\omega)={\cal A}_{z}(\vec{k},k_{z},\omega)e^{2}s\frac{g_{0}}{2} \omega_{c}\sum_{n,\sigma} \int \frac{dp_{z}}{2 \pi}v_{z}^{2}(p_{z})\int \frac{d\varepsilon}{4 \pi i}f_{\varepsilon}\left\{\left(G^{R}_{\sigma,n,p_{z},\varepsilon}\right)^{2}-\left(G^{A}_{\sigma,n,p_{z},\varepsilon}\right)^{2}\right\}.
\end{equation}

%Note : The diamagnetic current is always real : the corresponding conductivity is ima%ginary.

\subsection{Paramagnetic current}
The paramagnetic part of the averaged current density is given by

\begin{equation}
j^{P}_{j}(\vec{r},\tau)=-i e t s^{2}\sum_{\sigma}\left[G^{(1)}_{\sigma,j+1, j}(\vec{r},\tau,\vec{r},\tau)-G^{(1)}_{\sigma,j, j+1}(\vec{r},\tau,\vec{r},\tau) \right].
\end{equation}
We are looking for the Fourier transform of the current density 
\begin{equation}
j^{P}_{z}(\vec{k},k_{z}, \omega)=s \int\!\!\! \!\int \!\!d^{2}rd\tau e^{i \omega \tau-i\vec{k} \cdot \vec{r}} \sum_{j}e^{-i k_{z}sj}j_{j}^{P}(\vec{r},\tau).
\end{equation}
The summation over $j$ yields
\begin{eqnarray}
s \sum_{j}e^{-ik_{z}sj}\left(G^{(1)}_{j+1, j}-G^{(1)}_{j, j+1} \right)
%&=&
%s \sum_{j}
%e^{-ik_{z}sj} \int \! \!\!\!\int \frac{dp_{z}dp'_{z}}{(2\pi)^{2}}
%(e^{ip_{z}s(j+1)-ip'_{z}sj}-e^{ip_{z}sj-ip'_{z}s(j+1)})G^{(1)}_{p_{z},p'_{z}}
%\nonumber
%\\
%&=&\int\!\! \!\!\int\frac{dp_{z}dp'_{z}}{(2\pi)^{2}}\delta(p_{z}-p'_{z}-k_{z})
%(e^{ip_{z}s}-e^{-ip'_{z}s})G^{(1)}_{p_{z},p'_{z}} \nonumber \\
%&
=
%&
\int \frac{dp_{z}}{2\pi}(e^{ip_{z}s}-e^{-i(p_{z}-k_{z})s})G^{(1)}_{p_{z},p_{z}-k_{z}}.
\end{eqnarray}
By definition, we have
\begin{equation}
G^{(1)}_{p_{z},p'_{z}}(\vec{r},\tau,\vec{r},\tau)=\sum_{l,m}e^{-ip_{z}sl+ip'_{z}sm}G^{(1)}_{l,m}(\vec{r},\tau,\vec{r},\tau).
\end{equation}
Using (A26) and (A27), we obtain
\begin{equation}
G^{(1)}_{p_{z},p'_{z}}(\vec{r},\tau,\vec{r},\tau)=-\int \!\!\!\!\int\!\! d^{2}r''d\tau'' G^{0}_{p_{z}}(\vec{r},\vec{r}'',\tau-\tau'')ev_{z}(p'_{z})
{\cal A}_{z}(\vec{r''},p_{z}-p'_{z},\tau'')
G^{0}_{p'_{z}}(\vec{r}'',\vec{r},\tau''-\tau).
\end{equation}
Making the inverse Fourier series transformation with respect $\vec{r}$ and the Fourier series transformation with respect to $\tau$, we have
\begin{equation}
\int\!\!\! \!\int \!\!d^{2}rd\tau e^{i \omega \tau-i\vec{k} \cdot \vec{r}}G^{(1)}_{p_{z},p_{z}'}(\vec{r},\tau,\vec{r},\tau)=
-
\frac{g_{0}}{2} \omega_{c}
T \sum_{n,\omega_{\nu}}
G^{0}_{n,p_{z}}(\omega_{+})
ev_{z}(p'_{z})
{\cal A}_{z}(\vec{k},p_{z}-p'_{z},\omega)
G^{0}_{n,p'_{z}}(\omega_{-}),
\end{equation}
where $\omega_{\pm}=\omega_{\nu} \pm \omega/2$.
Collecting (A36), (A37) and (A40), we finally obtain

\begin{eqnarray}
j^{P}(\vec{k},k_{z}, \omega) = i{\cal A}_{z}(\vec{k},k_{z}, \omega)e^{2}ts^{2}\frac{g_{0}}{2} \omega_{c} \sum_{n,\sigma} \int \frac{dp_{z}}{2 \pi}\left(
e^{ip_{z}s}-e^{-i(p_{z}-k_{z})s} \right)v_{z}(p_{z}-k_{z}) \nonumber \times \\
\times
T \sum_{\omega_{\nu}}
G^{0}_{\sigma,n,p_{z}}(\omega_{+})
G^{0}_{\sigma,n,p_{z}-k_{z}}(\omega_{-}).
\end{eqnarray}
We are interested in the limit $k_{z} \ll p_{z}$, because $E_{z}$ is assumed to be uniform. Then,
\begin{equation}
j^{P}(\vec{k},k_{z}, \omega) = -{\cal A}_{z}(\vec{k},k_{z}, \omega)e^{2}s\frac{g_{0}}{2} \omega_{c}\sum_{n,\sigma}\int \frac{dp_{z}}{2 \pi} v_{z}^{2}(p_{z}) T \sum_{\omega_{\nu}}G^{0}_{\sigma,n,p_{z}}(\omega_{+})G^{0}_{\sigma,n,p_{z}}(\omega_{-}).
\end{equation}
Next, after transforming the summation over Matsubara frequencies into an integration over energy $\varepsilon$ and taking the analytic continuation~\cite{20}, we get

\begin{eqnarray}
T \sum_{\omega_{\nu}}&&
G^{0}_{p_{z}}(\omega_{+})
G^{0}_{p_{z}}(\omega_{-})=
T \sum_{\omega_{\nu}}G^{0}_{p_{z}}(\omega_{\nu})
G^{0}_{p_{z}}(\omega_{\nu}-\omega)\nonumber \\
&&
=
\int\!\!\! \frac{d \varepsilon}{4 \pi i}
\left\{
 \underbrace{
G^{R}_{p_{z}, \varepsilon} \left
(f_{\varepsilon}-f_{\varepsilon-\omega}\right)
G^{A}_{p_{z}, \varepsilon-\omega}}_{RA}+ 
\underbrace{G_{p_{z}, \varepsilon}^{R}
f_{\varepsilon-\omega}
G_{p_{z}, \varepsilon-\omega}^{R}
-G_{p_{z}, \varepsilon}^{A}f_{\varepsilon}G_{p_{z}, \varepsilon-\omega}^{A}}_{RR}
\right\}.
\end{eqnarray}

\subsection{Static conductivity}

At zero order of $\omega$,  the paramagnetic current density arising from the term $RR$ is cancelled with the diamagnetic current density (see Eq. (A34))
\begin{equation}
j_{k, \omega}^{P, RR}+j^{D}_{k, \omega}=0.
\end{equation}
At first order of $\omega$, a nonzero total current density arises from the terms $RA +RR$ of the paramagnetic current.
Using Eq. (A5), we find the static conductivity 
\begin{equation}
\sigma_{zz}=e^{2}s\frac{g_{0}}{2}\omega_{c} \sum_{n,\sigma}\int \frac{dp_{z}}{2 \pi}v^{2}_{z}(p_{z})\int \frac{d\varepsilon}{4\pi}\left(f'_{\varepsilon}G^{R}_{p_{z},\varepsilon}G^{A}_{p_{z},\varepsilon}
+2f_{\varepsilon}\Re \,\left( G^{R}_{p_{z},\varepsilon}G^{'R}_{p_{z},\varepsilon}\right)\right),
\end{equation}
where the prime means the derivative with respect to the energy $\varepsilon$.
Integrating by parts over the energy in the term $RR$ and using (A33), we obtain Eq. (2).

\end{document}